\documentclass[a4paper,aps,11pt,nofootinbib]{revtex4}
\usepackage{amsmath}
\usepackage{amssymb}
\usepackage{amsfonts}
\usepackage[dvipdfm]{graphicx}
\usepackage{color}
\usepackage{float}
\usepackage{comment}
\usepackage{ulem}
\usepackage{ascmac}
\usepackage{subfigure} 

\definecolor{DarkBlue}{rgb}{0,0,0.7}

\definecolor{DarkRed}{rgb}{0.65,0,0}

\definecolor{DarkGreen}{rgb}{0,0.3,0}

\definecolor{purple}{rgb}{0.7,0,0.7}

\begin{document}

\def\uwave{ }
\title{\large 
Solitonic Gravastars in a U(1) gauge Higgs model
}

\hfill{OCU-PHYS 576}

\hfill{~AP-GR 187}

\hfill{NITEP 153}


\author{\vspace{1cm}\large Tatsuya Ogawa}
\email{taogawa@omu.ac.jp}
\affiliation{
Osaka Central Advanced Mathematical Institute, Osaka Metropolitan University,
  Osaka 558-8585, Japan}
\author{\large Hideki Ishihara}
\email{h.ishihara@omu.ac.jp}
\affiliation{
Nambu Yoichiro Institute of Theoretical and Experimental Physics (NITEP),
Osaka Central Advanced Mathematical Institute, Osaka Metropolitan University,
  Osaka 558-8585, Japan
}

\begin{abstract}
\vspace{1cm}
We numerically obtain gravastar solutions as nontopological solitons 
in a system that consists of a U(1) gauge Higgs model 
with a complex scalar field and Einstein gravity.
The solitonic gravastar solutions are compact enough to have a photon sphere. 
\end{abstract}

\maketitle

\section{Introduction}

Black holes, mathematical solutions 
to the Einstein equations, 
are widely accepted as astrophysical objects.  
Recently, two important observations concerning to black holes in our universe 
have been reported. 
One is detection of gravitational waves from black hole binaries 
\cite{LIGOScientific:2016aoc, LIGOScientific:2021djp}, 
and the other is photographic evidences of black holes at the center of M87 and Sgr A* 
in our Galaxy \cite{EventHorizonTelescope:2019dse, EventHorizonTelescope:2022xnr}. 
However, observable phenomena for distant observers should occur outside the event horizon, 
or before the formation of event horizons. 
Therefore, verification of astrophysical object with event horizon is still an open question. 

As an alternative to black holes, a compact non-singular object, the so-called `gravastar', 
has been proposed to take quantum effects into account \cite{Mazur:2001fv, Mazur:2004fk}. 
The interior geometry of the gravastar is described by a de Sitter metric 
and the exterior is by a Schwarzshild metric, 
and these two regions are joined by a spherical shell with a finite thickness. 
The radius of the shell is smaller than the de Sitter horizon and larger than 
the Schwarzshild radius. Then, the gravastar has no horizon and no central singularity.

Nontopological solitons, on the other hand, have been studied as interesting astrophysical 
objects \cite{ Coleman:1985ki, Friedberg:1976me, Lee:1991ax, Shnir:2018yzp, 
Kusenko:1997si, Kusenko:2001vu, Enqvist:1997si, Kasuya:1999wu}.
In a U(1) gauge Higgs model coupled to a complex scalar field, 
various types of nontopological solitons 
are obtained \cite{Ishihara:2018rxg,Ishihara:2019gim,Ishihara:2021iag, 
Forgacs:2020vcy, Forgacs:2020sms}. 
In a type of nontopological solitons, called `potential balls' 
in ref.\cite{Ishihara:2021iag}, 
the vacuum energy of the Higgs scalar field is surrounded by a spherical shell, 
and the energy vanishes outside the shell. 
If the potential ball couple to the Einstein gravity, 
we would expect to obtain a solitonic gravastar solution.\footnote{
Nontopological soliton solutions with dust-like energy 
that couple to the Einstein gravity were obtained \cite{Endo:2022uhe}.}

\section{Model }

We consider the theory described by the action
\begin{align}
	S=\int &\sqrt{-g}d^4x \left(	\frac{R}{16\pi G}\right.
	-g^{\mu\nu}(D_{\mu}\psi)^{\ast}(D_{\nu}\psi)
\cr&
	-g^{\mu\nu}(D_{\mu}\phi)^{\ast}(D_{\nu}\phi)
 	-\frac{\lambda}{4}(\left|\phi \right|^2-\eta^2)^2-\mu \left|\psi \right|^2\left|\phi \right|^2
\cr
	&\left.	-\frac{1}{4}g^{\mu\nu}g^{\alpha\beta}F_{\mu\alpha}F_{\nu\beta}\right),
\label{eq:action}
\end{align}
where $R$ is the scalar curvature of a metric $g_{\mu\nu}$, $g$ denotes $\det(g_{\mu\nu})$, 
$G$ is the gravitational constant, 
$\psi$ and $\phi$ are complex scalar fields, 
and $F_{\mu\nu}:=\partial_{\mu}A_{\nu}-\partial_{\nu}A_{\mu}$ is the field strength of
a U(1) gauge field $A_\mu$, respectively.
The scalar field $\phi$ has the self-coupling term 
characterized by a constant $\lambda$ and
a symmetry breaking scale $\eta$, 
and the interaction term with $\psi$ characterized by a constant $\mu$. 
The both scalar fields interact with the gauge field through the gauge covariant derivative, 
$D_{\mu}:=\partial_{\mu}-ieA_{\mu}$ with a coupling constant $e$.

By varying the action \eqref{eq:action}, we obtain a coupled system of the field equations, 
\begin{align}
  &\frac{1}{\sqrt{-g}}D_{\mu}\left(\sqrt{-g}g^{\mu\nu}D_{\nu}\psi\right)
	-\mu \psi \left|\phi \right|^2=0,
\label{eq:ELequation_psi}  \\
  &\frac{1}{\sqrt{-g}}D_{\mu}\left(\sqrt{-g}g^{\mu\nu}D_{\nu}\phi\right)
		-\frac{\lambda}{2}\phi(\left|\phi \right|^2-\eta^2)-\mu \left|\psi \right|^2 \phi=0,
\label{eq:ELequation_phi} \\
  &\frac{1}{\sqrt{-g}}\partial_{\mu}(\sqrt{-g}F^{\mu\nu})=j^{\nu}_{\psi}+j^{\nu}_{\phi},
\label{eq:ELequation_A} \\
  &R_{\mu\nu}-\frac12Rg_{\mu\nu}=8\pi GT_{\mu\nu},
\label{eq:Einsteinequation}
\end{align}
where $j^{\mu}_{\psi}$ and $j^{\mu}_{\phi}$ are four-currents defined by
\begin{align}
	&j^{\mu}_{\psi}:=ie\left(\psi^{\ast}(D^{\mu}\psi)-(D^{\mu}\psi)^{\ast}\psi\right), 
\cr
	&j^{\mu}_{\phi}:=ie\left(\phi^{\ast}(D^{\mu}\phi)-(D^{\mu}\phi)^{\ast}\phi\right),
\label{eq:current_phi_psi}
\end{align}
$R_{\mu\nu}$ is the Ricci tensor, and $T_{\mu\nu}$ is the energy-momentum tensor defined by
\begin{align}
	T_{\mu\nu}
	&= 2(D_{\mu}\psi)^{\ast}(D_{\nu}\psi)
	-g_{\mu\nu}	(D_{\alpha}\psi)^{\ast}(D^{\alpha}\psi)
\cr
	&+2(D_{\mu}\phi)^{\ast}(D_{\nu}\phi)
	-g_{\mu\nu}(D_{\alpha}\phi)^{\ast}(D^{\alpha}\phi)
\cr
	&-g_{\mu\nu}\left( \frac{\lambda}{4}(|\phi|^2-\eta^2)^2
	 + \mu |\psi|^2|\phi|^2 \right)
\cr
	&+\left( F_{\mu\alpha}F_{\nu}^{~\alpha}
	-\frac{1}{4}g_{\mu\nu}F_{\alpha\beta}F^{\alpha\beta}\right).
\label{eq:T_munu}
\end{align}
The action \eqref{eq:action} is invariant under the transformations
\begin{align}
  	&\psi(x) \to \psi'(x)=e^{i(\chi(x)-\gamma)}\psi(x),
 \label{eq:psi_tr} \\
  	&\phi(x) \to \phi'(x)=e^{i(\chi(x)+\gamma)}\phi(x),
 \label{eq:phi_tr} \\
 	&A_{\mu}(x)\to A_{\mu}'(x)=A_{\mu}(x)+e^{-1}\partial_{\mu}\chi(x),
 \label{eq:A_tr}
\end{align}
and we have conservation equations of the currents:
\begin{align}
	\partial_\mu(\sqrt{-g}~ j^{\mu}_{\psi})=0 
	~\mbox{and}~ \partial_\mu (\sqrt{-g}~j^{\mu}_{\phi})=0. 
\end{align}
We assume a static and spherically symmetric metric, 
\begin{align}
  ds^2
  =&-\sigma(r)^2\left(1-\frac{2m(r)}{r}\right)dt^2+\left(1-\frac{2m(r)}{r}\right)^{-1}dr^2
	+r^2d\theta^2+r^2\sin^2\theta d\varphi^2,
\label{eq:metric}
\end{align}
and spherically symmetric fields in the form 
\begin{align}
  	\psi=e^{-i\omega t}u(r), \quad \phi=e^{-i\bar\omega t}f(r), \quad A_{\mu}dx^{\mu}=A_t(r)dt, 
  \label{eq:matter_ansatz}
\end{align}
where $\sigma(r)$, $m(r), u(r)$, $f(r)$, and  $A_t(r)$, are functions of $r$, 
and parameters $\omega$ and $\bar\omega$ are constants. 
Total charges of $\psi$ and $\phi$ on the $t=const.$ slices defined by
\begin{align}
	Q_{\psi}:=\int d^3x \sqrt{-g}~\rho_{\psi} , \quad
	Q_{\phi}:=\int d^3x \sqrt{-g}~\rho_{\phi},
\label{eq:charge_phi_psi}
\end{align}
are conserved, respectively, 
where $\rho_{\psi}:=j^t_{\psi}$ and $\rho_{\phi}:=j^t_{\phi}$.

By using the gauge transformation \eqref{eq:psi_tr}-\eqref{eq:A_tr}, 
we can fix the variables as 
\begin{align}
&\psi(t,r) \to e^{i\Omega t}u(r), \quad     \phi(t,r) \to f(r), \quad 
\cr
&	A_t(r) \to \alpha(r):= A_t(r)+e^{-1}\bar\omega,
\label{eq:matter_variable}
\end{align}
where $\Omega:=\bar\omega-\omega$.
Substituting \eqref{eq:metric} and \eqref{eq:matter_variable}
into \eqref{eq:ELequation_psi}-\eqref{eq:Einsteinequation},
we obtain a system of coupled ordinary differential equations to be solved 	in the form:
\begin{align}
  &u''+\left(\frac{2}{r}\left(1+\frac{m-rm'}{r-2m}\right)+\frac{\sigma'}{\sigma}\right)u'
	+\left(1-\frac{2m}{r}\right)^{-1}\left(\frac{(e\alpha-\Omega)^2u}{\sigma^2(1-2m/r)}
	-\mu f^2u\right)=0,
 \label{eq:equation_u}
\\
 &f''+\left(\frac{2}{r}\left(1+\frac{m-rm'}{r-2m}\right)+\frac{\sigma'}{\sigma}\right)f'
\cr& \hspace{2cm}
	+\left(1-\frac{2m}{r}\right)^{-1}\left(\frac{e^2f\alpha^2}{\sigma^2(1-2m/r)}
	-\frac{\lambda}{2}f(f^2-\eta^2)-\mu fu^2\right)=0,
 \label{eq:equation_f}
\\
  &\alpha''+\left(\frac{2}{r}-\frac{\sigma'}{\sigma}\right)\alpha'
	+\left(1-\frac{2m}{r}\right)^{-1}\left(-2e^2f^2\alpha -2e(e\alpha-\Omega) u^2\right)=0,
\label{eq:equation_alpha}
\\
  &\frac{2m'}{r^2}-8\pi G\biggl[
	\frac{e^2f^2\alpha^2+(e\alpha-\Omega)^2u^2}{\sigma^2(1-2m/r)}
	+\left(1-\frac{2m}{r}\right)\left(\left(\frac{df}{dr}\right)^2
	+\left(\frac{du}{dr}\right)^2\right)
\cr
  &\hspace{2cm}
 +\frac{\lambda}{4}(f^2-\eta^2)^2+\mu f^2u^2+\frac{1}{2\sigma^2}\left(\frac{d\alpha}{dr}\right)^2 \biggr]=0,
   \label{eq:equation_Rtt}
\\
	&\frac{(1-2m/r)\sigma'}{r\sigma}-8\pi G\left[
	\frac{e^2f^2\alpha^2+(e\alpha-\Omega)^2u^2}{\sigma^2(1-2m/r)}
\right.\cr&\left.\hspace{4cm}
	+\left(1-\frac{2m}{r}\right)\left(\left(\frac{df}{dr}\right)^2
	+\left(\frac{du}{dr}\right)^2\right)\right]=0.
\label{eq:equation_RttRrr}
\end{align}
where prime denotes the derivative with respect to $r$.

We require that all fields at the origin be regular, 
and that the scalar fields and the gauge field be localized in a finite region,
then we impose
\begin{align}
	\frac{d\sigma}{dr}=0, \quad m=0, \quad \frac{du}{dr}=0, \quad
	\frac{df}{dr}=0, \quad \frac{d\alpha}{dr}=0, \qquad \text{at $r=0$, }
\label{eq:BC_origin}
\end{align}
and 
\begin{align}
	u=0, \quad f=\eta, \quad \alpha=0, \qquad \text{at spatial infinity.}
\label{eq:BC_infinity_matter}
\end{align}
On these assumptions, 
the geometry should be described by a Schwarzschild metric in a far region, 
namely, we can impose
\begin{align}
	\sigma=1, \quad m=m_{\infty}=\text{\it const.}, \qquad \text{at spatial infinity.}
\label{eq:BC_infinity_metric}
\end{align}

\section{Solitonic Gravastar solutions }

\begin{figure}[H]
\centering
\includegraphics[width=6cm]{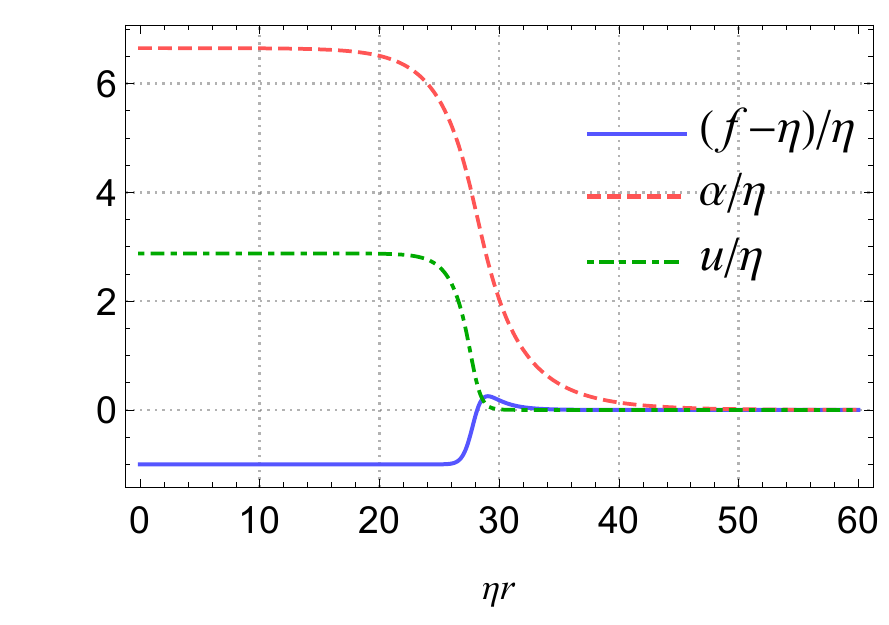}~
\includegraphics[width=6cm]{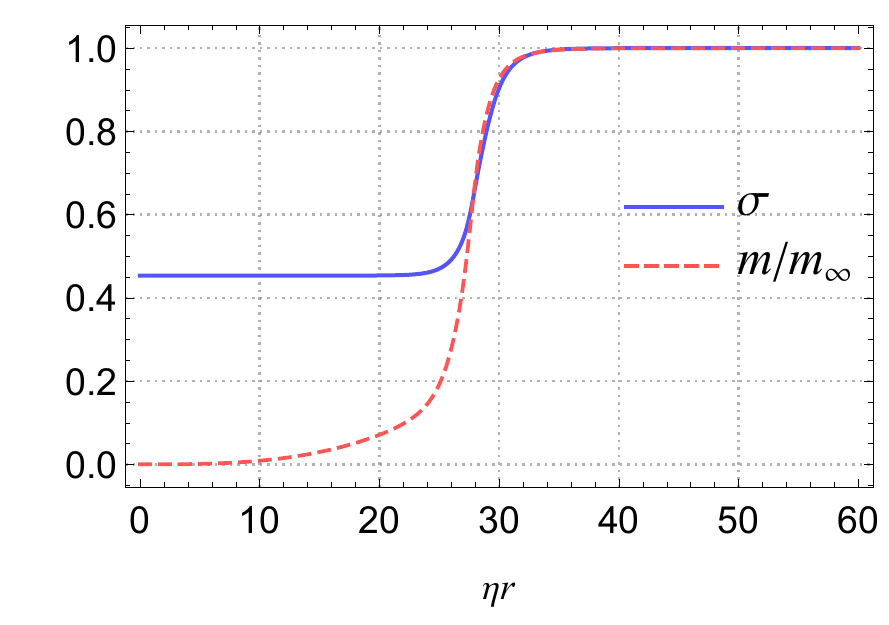}
\caption{
Field configurations of the numerical solution for the parameter $\Omega/\eta=0.665$. 
The scalar fields $u, f$ and the gauge field $\alpha$ are plotted in the left panel, 
and the metric components $\sigma$ and $m$ are plotted in the right panel.
At the origin, $r=0$, it is found that $\alpha=\Omega/e$ and $f=0$. 
The mass at infinity is obtained numerically as $m_\infty= 11.94 \eta^{-1}$. 
}
\label{fig:comfigulation_G10e4_Omega0p665}
\end{figure}

We fix the coupling constants 
as $e=0.1$, $\mu=1.4$, and $\lambda=1.0$, and 
we set the symmetry breaking scale $\eta = 10^{-2} M_P$, for an example
\footnote{
For the set of parameters, the potential balls are found as solutions 
in the case that the gravity is decoupled \cite{Ishihara:2021iag}.
}. 
In Fig.\ref{fig:comfigulation_G10e4_Omega0p665}, the field variables of a numerical solution 
are shown as functions of $r$. 
The matter variables $u, f$, and $\alpha$ change quickly in a layer 
of thickness $\Delta r \sim 10 \eta^{-1}$ around radius $r=r_{\rm sl}\sim 28\eta^{-1}$. 
We call the layer the surface layer. 

\begin{figure}[H]
\centering
\includegraphics[width=6cm]{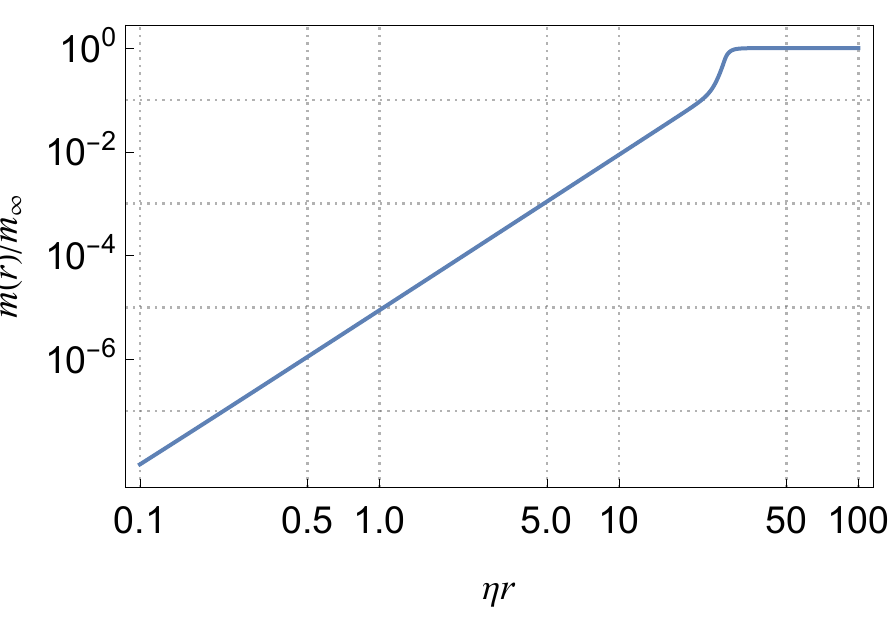}
\caption{The mass variable $m/m_\infty$
 is plotted as a function of $r$ on a log-log scale. 
}
\label{fig:loglogm_comfigulation_G10e4_Omega0p665}
\end{figure}

Outside the radius $r_{\rm sl}$, matter variables decay to the values for the symmetry-breaking vacuum,
namely the fields are excited in the compact region inside the radius. 
The fact that the metric functions $\sigma=1$ and $m=m_\infty=const.$ means 
the metric exhibits the Schwarzschild metric: 
\begin{align}
  ds^2=&-\left(1-\frac{2m_{\infty}}{r}\right)dt^2+\left(1-\frac{2m_{\infty}}{r}\right)^{-1}dr^2
		+r^2(d\theta^2+\sin^2\theta d\varphi^2), 
 \label{eq:effective_metric2}
\end{align}
where the value of $m_\infty$ is obtained numerically as $11.94 \eta^{-1}$.

Inside the radius $r_{\rm sl}$, we see that 
$u= const.$, $f=0$, $\alpha=\Omega/e$, 
then only the potential term of $\phi$ contributes to the energy-momentum tensor as 

\begin{align}
	T^t_t=T^r_r=T^\theta_\theta=T^\varphi_\varphi=-\frac{\lambda}4 \eta^4. 
\end{align}
Using the log-log plot of $m(r)$ 
in Fig.\ref{fig:loglogm_comfigulation_G10e4_Omega0p665}, 
we see that 
\begin{align}
  m(r)=\frac{\Lambda}{6} r^3, 
  \label{eq:m_inside}
\end{align}
where the value of $\Lambda$ is given by
\begin{align}
\Lambda=8\pi G~\frac{\lambda}{4} \eta^4 \sim 6.3\times 10^{-4}\eta^2. 
\end{align}
Furthermore, since $\sigma$ takes a constant, say $\sigma_0$, 
the geometry is described by the de Sitter metric given by
\begin{align}
  ds^2=&-\left(1-\frac{\Lambda}{3} r^2\right)d\tilde{t}^2
	+\left(1-\frac{\Lambda}{3} r^2\right)^{-1}dr^2
	+r^2 (d\theta^2+\sin^2\theta d\varphi^2) ,
 \label{eq:effective_metric1}
\end{align}
where $\tilde{t}:=\sigma_0 t$. 

The surface layer 
connects the de Sitter inner region and the Schwarzschild outer region. 
The de Sitter horizon radius, $r_{\rm dS}$, and the Schwarzschild radius, $r_{\rm Sch}$, 
of the numerical solution is estimated as 
\begin{align}
	r_{\rm dS} = \sqrt{\frac{3}{\Lambda}}
	\sim 0.7 \times 10^2 \eta^{-1},\quad
	r_{\rm Sch} = 2m_\infty \sim 24 \eta^{-1}, 
\end{align}
and therefore we have $r_{\rm Sch} <r_{\rm sl} <r_{\rm dS}$. 
The nontopological soliton solution describes the gravastar.

\begin{figure}[H]
\centering
\includegraphics[width=6cm]{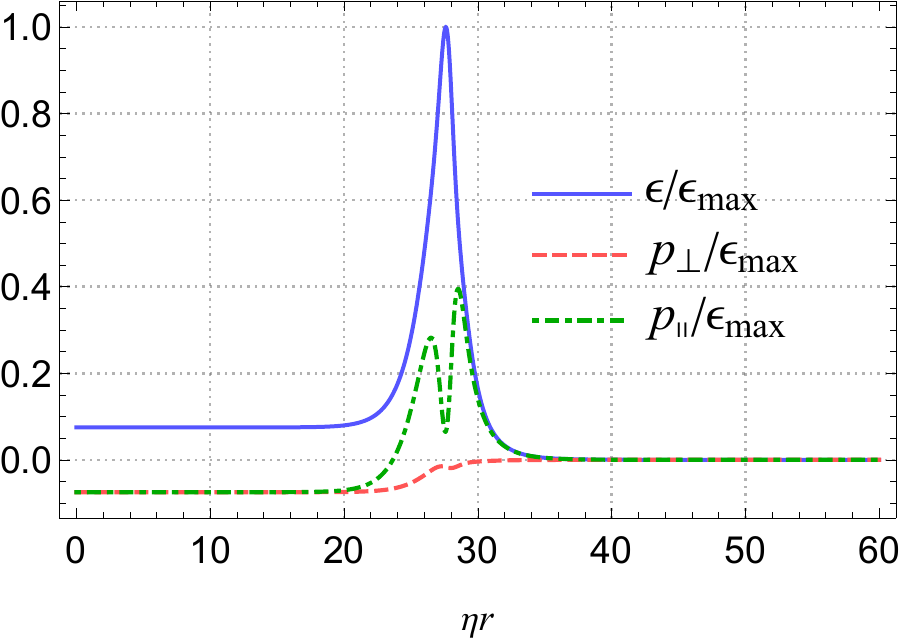}
\caption{
Energy density $\epsilon$, tangential pressure $p_\parallel$, 
and radial pressure $p_\perp$ are shown as functions of $r$.  
The pressure components are normalized by the maximum value of $\epsilon$.
}
\label{fig:MassPressure_G10e4}
\end{figure}

In Fig.\ref{fig:MassPressure_G10e4} we show energy density  $\epsilon=-T^t_t$, 
radial pressure $p_\perp=T^r_r$, and tangential pressure $p_{\parallel}=T^\theta_\theta=T^\varphi_\varphi$ 
for the numerical solution as functions of $r$.
The surface layer has the structure within its thickness 
given by the Compton length of the gauge field $\sim (e \eta)^{-1}$. 
The energy density $\epsilon$ has a peak, 
and $p_\parallel$ has two peaks with almost $1/3 \sim 2/5$ of the peak hight 
of $\epsilon$, while $p_\perp$ is almost zero. 

\begin{figure}[H]
\centering
\includegraphics[width=6cm]{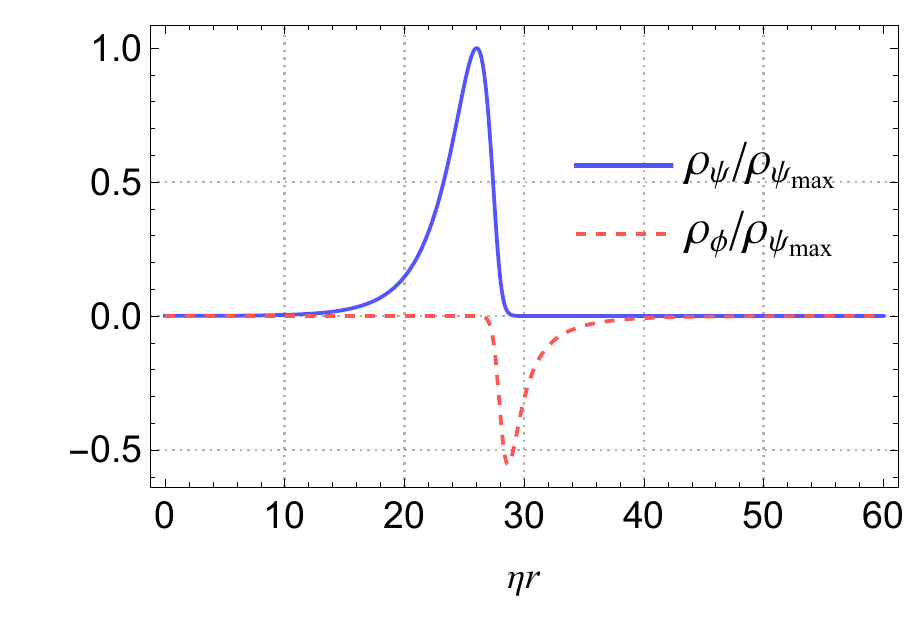}~~~~
\includegraphics[width=6cm]{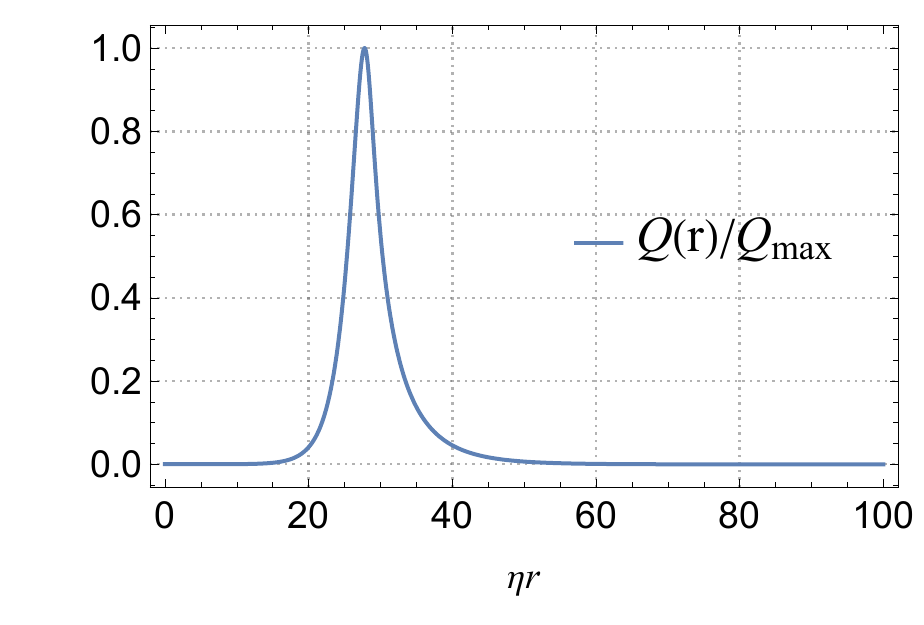}~~~~
\caption{The charge densities of the complex scalar fields, $\rho_{\psi}$ and $\rho_{\phi}$, 
normalized by the maximum value of $\rho_{\psi}$, $\rho_{\psi_{max}}=0.851\eta^{-3}$, 
are shown in the left panel. 
Total charge included within radius $r$, $Q(r)$, is shown in the right panel, 
where $Q_\text{max}=1.44\times 10^4$. 
}
\label{fig:Chargedensity_G10e4}
\end{figure}

We show the charge densities $\rho_{\psi}$ and $\rho_{\phi}$  
in Fig.\ref{fig:Chargedensity_G10e4} as functions of $r$. 
The positive $\rho_{\psi}$ is induced on the inner-side surface of the surface layer, 
and the negative $\rho_{\phi}$ on the outer-side (see the left panel). 
Namely, an electric double layer emerges at the surface layer. 
The total charge 
contained inside a radius $r$, shown in the right panel, 
decays quickly outside of the surface radius, 
namely, the charge is screened for a distant observer. 
Therefore, the radial electric field appears in the electric double layer. 
Owing to this charge screening effect, the geometry of outside is given 
by the Schwarzschild metric instead of Reissner-Nordstr\"om one.

For a numerical solution, 
we define the surface radius of the solitonic gravastar, say $r_{\rm gs}$, by
\begin{align}
	m(r_{\rm gs}):=0.99~ m_{\infty}, 
\label{eq:numerical_R}
\end{align}
namely, $99\%$ of total mass of the solitonic gravastar is included within the radius $r_{\rm gs}$.
For the numerical solution shown in Fig.\ref{fig:comfigulation_G10e4_Omega0p665}, 
we estimate $r_{\rm gs}\sim 33.2 \eta^{-1}$.
By the numerical values of 
$m_\infty$ 
and 
$r_{\rm gs}$, 
we estimate the compactness as
\begin{align}
 	C:=\frac{2 m_\infty }{r_{\rm gs}} \sim 0.718 \ge 2/3, 
  \label{eq:compactness}
\end{align}
then the solitonic gravastar is compact so that it has the photon sphere.

\section{Summary}

We have studied numerically the coupled system of a U(1) gauge Higgs model 
with a matter complex scalar field and Einstein gravity, 
which is characterized by a set of parameters: 
coupling constants and a symmetry-breaking scale. 
For a choice of the parameters, we have found the solitonic gravastar solutions. 
Each solution has an internal de Sitter geometry in the symmetric vacuum 
with the potential energy of the Higgs scalar field, 
and an external Schwarzschild geometry in the symmetry-breaking vacuum.  
These regions are joined by a spherical surface layer with a finite thickness 
that has nonvanishing tangential pressure. 
Within the thickness of the surface layer, an electric double layer 
is produced by the two complex scalar fields, 
and the total charge is screened for a distant observer. 
For the set of parameters used in this paper, 
the solitonic gravastar obtained is compact enough to have a photon sphere. 
Then, it is a compact regular object without the event horizon  
as an alternative to a black hole.   

For the numerical solutions, 
the total gravitational mass $M_G = m_\infty/G$ is of the order of 
$10^3$ times the Planck mass, which is much smaller than the astrophysical scale.  
The surface layer with the thickness about $1/3 $ times the radius of 
the solitonic gravastar has the internal structure. 
These are different properties from original gravastars, a final state of gravitational 
collapsing astrophysical objects, where solutions are 
constructed by using a thin shell approximation \cite{Mazur:2001fv, Mazur:2004fk}. 
However, as seen in the previous work \cite{Endo:2022uhe}, 
the total mass, surface radius, and thickness of the surface layer of the 
numerical solutions would depend on the model parameters. 
Therefore, it is interesting to clarify 
whether the solitonic gravastar can have astrophysical mass scale, 
and thickness of the surface layer becomes much smaller than its radius. 

There are important and interesting works on gravastar solutions: 
the stability of the solutions \cite{Visser:2003ge}, 
the behavior of null geodesics around the photon sphere \cite{Sakai:2014pga}, 
gravitational wave emission \cite{ Pani:2009ss},  
and Hawking radiation \cite{Nakao:2022ygj}. 
These issues are addressed using thin shell approximations. 
We aim to study these problems using solutions in U(1) gauge Higgs models 
as next works. 
Furthermore,  it would be interesting to investigate 
whether the solitonic gravastar 
solutions are a possible final state for the gravitational collapse of the system.

\section*{Acknowledgements}
We would like to thank K.-i. Nakao and H.Yoshino for valuable discussions and comments. 
This work was partly supported by Osaka Central Advanced Mathematical 
Institute: MEXT Joint Usage/Research Center on Mathematics and Theoretical
Physics JPMXP0619217849.

\end{document}